\documentclass[conference]{IEEEtran}
\IEEEoverridecommandlockouts

\usepackage{cite}
\usepackage{amsmath,amssymb,amsfonts}
\usepackage{algorithmic}
\usepackage{graphicx}
\usepackage{textcomp}
\usepackage{xcolor}

\usepackage{booktabs}       
\usepackage{amssymb} 
\usepackage{multicol}
\usepackage{multirow}
\usepackage{pifont}

\def\BibTeX{{\rm B\kern-.05em{\sc i\kern-.025em b}\kern-.08em
    T\kern-.1667em\lower.7ex\hbox{E}\kern-.125emX}}
\begin{document}

\title{DapPep: Domain Adaptive Peptide-agnostic Learning for Universal T-cell Receptor-antigen Binding Affinity Prediction
}

\author{\IEEEauthorblockN{1\textsuperscript{st} Jiangbin Zheng}
\IEEEauthorblockA{\textit{Zhejiang University} \\
\textit{Westlake University}\\
Hangzhou, China \\
zhengjiangbin@westlake.edu.cn}
\and
\IEEEauthorblockN{2\textsuperscript{nd} Qianhui Xu}
\IEEEauthorblockA{\textit{Fudan University} \\
Shanghai, China \\
qhxu22@m.fudan.edu.cn}
\and
\IEEEauthorblockN{3\textsuperscript{rd} Ruichen Xia}
\IEEEauthorblockA{\textit{Zhejiang University} \\
Hangzhou, China \\
rcxia@zju.edu.cn}
\and
\IEEEauthorblockN{4\textsuperscript{th} Stan Z. Li*}
\IEEEauthorblockA{\textit{Westlake University} \\
Hangzhou, China \\
stan.zq.li@westlake.edu.cn\\
*Corresponding Author}
}

\maketitle

\begin{abstract}
Identifying T-cell receptors (TCRs) that interact with antigenic peptides provides the technical basis for developing vaccines and immunotherapies. The emergent deep learning methods excel at learning antigen binding patterns from known TCRs but struggle with novel or sparsely represented antigens. However, binding specificity for unseen antigens or exogenous peptides is critical. We introduce a domain-adaptive peptide-agnostic learning framework DapPep for universal TCR-antigen binding affinity prediction to address this challenge. The lightweight self-attention architecture combines a pre-trained protein language model with an inner-loop self-supervised regime to enable robust TCR-peptide representations. Extensive experiments on various benchmarks demonstrate that DapPep consistently outperforms existing tools, showcasing robust generalization capability, especially for data-scarce settings and unseen peptides. Moreover, DapPep proves effective in challenging clinical tasks such as sorting reactive T cells in tumor neoantigen therapy and identifying key positions in 3D structures. 
\end{abstract}


\begin{figure*}[t!]
    \centering
    \includegraphics[width=0.99\linewidth]{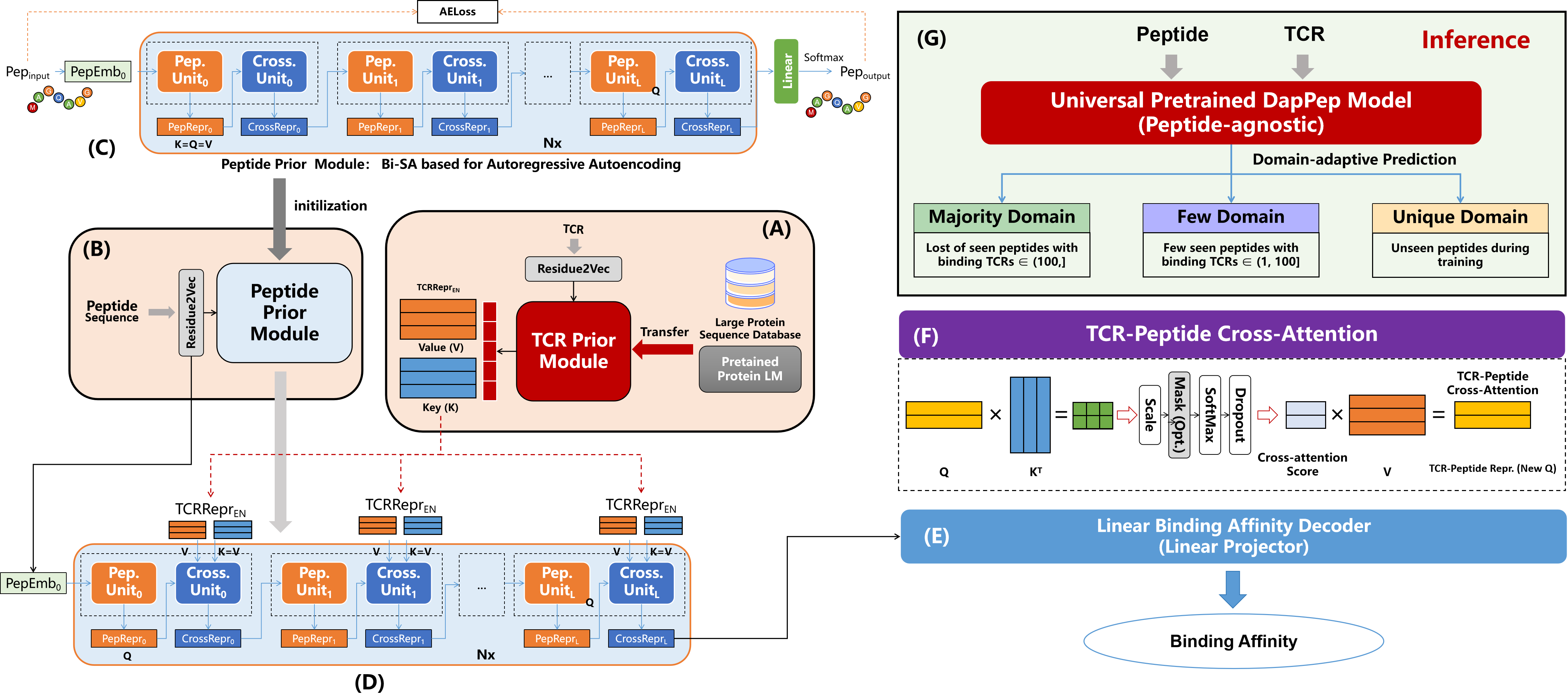}
    \vspace{-1em}
    \caption{
    Proposed DapPep for TCR-peptide binding affinity prediction. The training pipeline is divided into two stages: Stage 1 involves initializing the TCR representation module (A) and pre-training the peptide prior module (B, C), while Stage 2 entails the pre-trained modules transferred from Stage 1 to optimize the overall framework (D).
    (E). Linear binding affinity decoder.
    (F). TCR-peptide cross-attention module.
    (G). Inference regimes for different data settings.
    }
    \label{fig:1}
    \vspace{-0.5em}
\end{figure*}

\section{Introduction}\label{sec:intro}

Computational recognition of T-cell receptor (TCR)-peptide complexes is crucial for understanding tumors, autoimmune, and viral infectious diseases \cite{schumacher2015neoantigens,linette2017neoantigen,ott2017immunogenic}, where antigenic and viral peptides are presented by MHC-I. TCRs on T cells recognize these complexes, triggering an immune response. As key molecules in acquired immunity, TCRs exhibit complex diversity due to genetic recombination and evolutionary screening. Predicting TCR-peptide binding is a fundamental challenge in computational immunology, offering opportunities for vaccine development and immunotherapy. While traditional experimental methods \cite {altman1996phenotypic,zhang2018high,kula2019t} exist to detect TCR-MHC interactions, they are often time-consuming and technically difficult.

With the rise of deep learning, many approaches have emerged for analyzing TCR patterns and predicting TCR-peptide binding affinity, categorized as follows:
1) Quantitative similarity measurement methods \cite{dash2017quantifiable,sidhom2021deeptcr,zhang2021giana,zhang2020investigation,glanville2017identifying,huang2020analyzing,dvorkin2021autoencoder}, which cluster TCRs and decipher antigen-specific binding patterns, but are not directly applicable for TCR-peptide binding prediction.
2) Peptide-specific TCR binding prediction models \cite{jokinen2021predicting,gielis2019detection,montemurro2021nettcr}, which are limited to specific peptides, restricting their utility.
3) Peptide-agnostic TCR-peptide binding prediction models, such as pMTnet\cite{lu2021deep}, DLpTCR\cite{xu2021dlptcr}, ERGO2\cite{springer2020prediction} and TITAN\cite{weber2021titan}, which can handle a broader range of TCR-peptide pairs but struggle with generalizing to unseen peptides or those with limited TCR interactions, posing challenges for immunotherapy.
4) PanPep \cite{gao2023pan}, a recent study, improves generalization through various settings (majority, few-shot, and zero-shot). However, it relies on fine-tuning in few-shot cases and struggles with unseen peptides in zero-shot settings. PanPep is a pseudo-peptide-agnostic method combining peptide-agnostic and peptide-specific models, and its effectiveness depends on experience and further fine-tuning.

In addition to the limitations of tool design, the diversity and complexity of the data remain the main reasons for the accurate identification.
On the one hand, TCRs exhibit a high degree of diversity, making it difficult to generalize existing computational tools. On the other hand, known peptide-TCR pairing data obey a long-tailed distribution \cite{gao2023pan}, resulting in a severely uneven distribution, in which a few peptides combine lots of known TCR binding data, but most peptides record only a few known TCR binding information.
In this case, the conventional supervised paradigm leads to learning binding patterns in the majority setting and is difficult to generalize to the few/zero-shot settings.

To tackle these challenges, we propose a novel domain-adaptive peptide learning framework, \textbf{DapPep}, for TCR-peptide binding affinity prediction, as illustrated in Figure~\ref{fig:1}. DapPep is a universal peptide-agnostic model capable of adapting to various data settings (e.g., majority, few-shot, and zero-shot) without bias. Experimental results across multiple benchmarks demonstrate DapPep's strong generalization capabilities, significantly outperforming existing methods, particularly for unseen peptides like exogenous or neoantigens. DapPep has also been successfully applied to complex clinical tasks, yielding promising results in line with expectations for challenging treatments.

The main contributions are as follows:

$\bullet$ DapPep is a novel tool for peptide-agnostic TCR-peptide binding affinity prediction that can be well-generalized for exogenous or neonatal antigen identification and clinical applications with high throughput and efficiency.

$\bullet$ The interpretable approaches introduce scientific benchmarks to further reveal the nature of TCR-peptide interactions.

\section{Methods}\label{sec:methods}

Peptide-agnostic DapPep consists of three key modules: TCR representations (\text{$T_{Repr}$}) module in Figure~\ref{fig:1} (A),  peptide representations (\text{$P_{Repr}$}) module in (B), and TCR-peptide representations (\text{$TP_{Repr}$}) module in  (D). Following the \text{$TP_{Repr}$} module is a linear decoder in Figure~\ref{fig:1} (E) for scoring binding affinity. 
For simplicity and efficiency, all modules are powered by multi-head self-attention layers. 
The training is divided into two stages: Stage 1 is the initialization of the \text{$T_{Repr}$} module and the pre-training of the \text{$TP_{Repr}$}. Since TCRs are special proteins, the \text{$T_{Repr}$} module can be initialized with parameters derived directly from off-the-shelf powerful protein language models.
The pre-trained \text{$TP_{Repr}$} module can be viewed as a sequence-to-sequence machine-translation task based on an asymmetric auto-encoder (asyAE) network.
Stage 2 is to transfer the pre-trained \text{$T_{Repr}$} and \text{$TP_{Repr}$} modules to the binding affinity learning pipeline.

\textbf{TCR Representation Module.}
The \text{$T_{Repr}$} focuses on TCR sequence representations. To leverage the potential rich information contained in TCRs, we initialize the \text{$T_{Repr}$} module with the existing pre-trained protein language model ESM-2 \cite{lin2022language}. Since ESM-2 is trained on large-scale protein data, transfer learning enhances effectively the capability of the \text{$T_{Repr}$} to capture the diverse and information-dense nature of TCRs.

\textbf{Peptide Representation Module.}
The \text{$P_{Repr}$} module is responsible for encoding the features of peptide sequences. 
Compared to TCRs, most peptides are short in length (typically less than 10), quite limited in variety, and monotonous in structure. Hence, to effectively represent the peptides, we opt for shallow self-attention layers to capture the relevant features.
This concise structure learns the peptide representation directly in an end-to-end manner, which makes it easier to fit the binding pattern of the TCR to the peptide.

\textbf{TCR-Peptide Cross-attention Module.}
The cross-attention \text{$TP_{Repr}$} module is at the heart of DapPep, enabling it to overcome the pair-sensitive limitations. By employing a cross-attention mechanism, the module helps to delve into the interdependencies between TCR and peptide residues at a finer level. 
Specifically, following the vanilla self-attention mechanism \cite{vaswani2017attention}, the $key$ and $value$ input to the cross-attention \text{$TP_{Repr}$} module are both from the TCR features of \text{$T_{Repr}$} module, and $query$ is from the peptide features of \text{$P_{Repr}$} module, as shown in Figure~\ref{fig:1}(F).

\subsection{Inner-loop TCR-Peptide Module Pre-training}
Instead of transfer learning, the \text{$TP_{Repr}$} module embraces the challenge of reconstructing peptide sequences through an asyAE framework.
On the one hand, there is no specialized large-scale pre-trained language model for peptides, possibly due to the limited number of trainable peptides and the instability of the peptide structure. On the other hand, it is unnecessary to use peptides extensively because of their short and structurally simple sequences.
By training the module to reconstruct peptides, it learns to capture the essential characteristics and unique patterns inherent in the peptide sequences. This unsupervised learning approach enhances the module's ability to represent peptides accurately, setting the stage for accurate binding affinity predictions.
Additionally, the pre-training process simulates the interaction between sequences to initialize the parameters of the cross-attention layers.

\textbf{Inner-loop asyAE Architecture.}
As depicted in Figure~\ref{fig:1}(C), both the input and output being peptides, the multi-head cross-attention network in the \text{$TP_{Repr}$} module degrades to self-attention layers. Consequently, the \text{$TP_{Repr}$} module only functions as the encoder for peptides (different from the \text{$P_{Repr}$} module in the binding affinity prediction pipeline).
Hence, the \text{$TP_{Repr}$} module receives the peptide sequence features (Word2Vec) as $key$, $query$, and $value$. This is also different from the input of the \text{$TP_{Repr}$} in the binding affinity prediction pipeline.
Acting as an encoder, it leverages the learned representations from the TCR and peptide modules to capture the intricate interplay between TCR and peptide residues. 
The peptide decoder, followed by the cross-attention encoder, essentially a non-autoregressive linear layer, aims to generate peptide sequences. The encoder-decoder architecture forms an asyAE model, as the model's objective is to reconstruct the peptide sequences. Shared amino acid embeddings and position encodings between the encoder and decoder, the asyAE architecture learns to preserve important contextual and positional features during the reconstruction process. 
To further enhance the learning process, the peptide decoder utilizes a lower triangular mask mechanism during training, which ensures that asyAE model is capable of effectively reconstructing the native sequences, leading to a more comprehensive understanding of the underlying sequence semantics.

\textbf{Pre-training Objective.}
Using only the peptide sequences from the training set of TCR-peptide pairs, we pre-train the proposed asyAE framework~\cite{zheng2023cvt,zheng2024metaenzyme,zheng2024progressive,zheng2023mmdesign,zheng2023lightweight,zheng2022using,zheng2021enhancing,zheng2020improved,zheng2022leveraging,kamal2019technical,tong2020document,xia2024understanding,huang2024protein,hu2023learning,hu2022protein,huang2023data,xia2024discognn,wu2020fuzzy}, as shown in Figure~\ref{fig:1}(C). This enables the asyAE to learn contextual semantic knowledge, while the cross-entropy loss is used to fit the probability distribution of the generated sequences.
Formally, we assume that an input peptide sequence is denoted as $S_{in}=\{s_1, s_2, \cdots, s_n\}$ with $n$ amino acids, and the recovered TCR sequence of the decoder is denoted as $S_{out}=\{s'_1, s'_2, \cdots, s'_n\}$. The corresponding probability distribution is denoted as $S_\text{logits}$. The reference native sequence is $S_{native} = S_{in} = \{s_1, s_2, \cdots, s_n\}$. The asyAE pre-training as:
\begin{equation}
S_\text{logits} = \textrm{PepDecoder}(\textrm{\text{$TP_{Repr}$}}(\textrm{Residue2Vec}(S_{in}))), 
\end{equation}
where $Residue2Vec$ represents the Word2Vec for amino acids, and $PepDecoder$ represents the peptide decoder.

The sequence recovery cross-entropy loss is defined as:
\begin{equation}
\mathcal{L}_{\textrm{seqCE}} = \textrm{CE} \big(\textrm{Softmax} (S_\textrm{logits}), S_\textrm{native}) \big).
\end{equation}


\subsection{Binding Affinity Prediction Pipeline}
As shown in Figure~\ref{fig:1}(F), the binding affinity learning stage aims to predict the binding scores for TCR-peptide pairs. The output features of cross-attention \text{$TP_{Repr}$} module are fed into the binding decoder module, which consists of mean pooling layers and simple linear layers. The binding decoder then passes the encoded features through a sigmoid function to produce a continuous binding affinity score ranging from 0 to 1, which reflects the probability of binding between a TCR-peptide pair. Higher affinity scores indicate a stronger likelihood of clone expansion for the TCR in response to the peptide. 
The binding affinity training target guides the model to learn and capture the complex binding patterns between TCRs and peptides, enabling it to accurately predict the affinity between unknown TCR-peptide pairs.
During the training phase, a mean squared error (MSE), i.e., squared L2 norm, is employed to calculate the loss criterion.

Formally, given a TCR-peptide pair, let $T=\{x_1, x_2, \cdots, x_n \}$ denote the input TCR sequence and $P=\{x_1, x_2, \cdots, x_m \}$ denote the input Peptide sequence for DapPep, where $n$ and $m$ is the length of the sequences. The corresponding reference binding affinity score is represented as $\text{Score}_{target} \in [0,1]$. 
The TCR sequence $T$ is first transformed into its dense features $\text{feat}_{tcr}$ via \text{$T_{Repr}$} module, and the peptide sequence $P$ is simultaneously transformed into its dense features $\text{feat}_{pep}$ via \text{$P_{Repr}$} module. These features are then used as inputs for the TCR-Pep cross-attention module, which generates a combined representation features $\text{feat}_{pep}$ capturing the interdependencies among TCR-peptide residues. Finally, the binding decoder $\text{BindDecoder}$ takes the $\text{feat}_{pep}$ as input and predicts the binding affinity score $\text{Score}_{pred}$. 
The entire pipeline can be expressed as:
\begin{equation}
\text{Score}_{pred} = \textrm{BindDecoder}(\text{feat}_{tcr}, \text{feat}_{pep}) \in [0,1].
\end{equation}

The final MSE loss is calculated as follows:
\begin{equation}
\mathcal{L}_{\textrm{DapPep}}= \textrm{MSE} (\text{Score}_{pred}, \text{Score}_{target})).
\end{equation}

\section{Experiments}

\begin{figure*}[htp]
    \centering
    \includegraphics[width=0.86\linewidth]{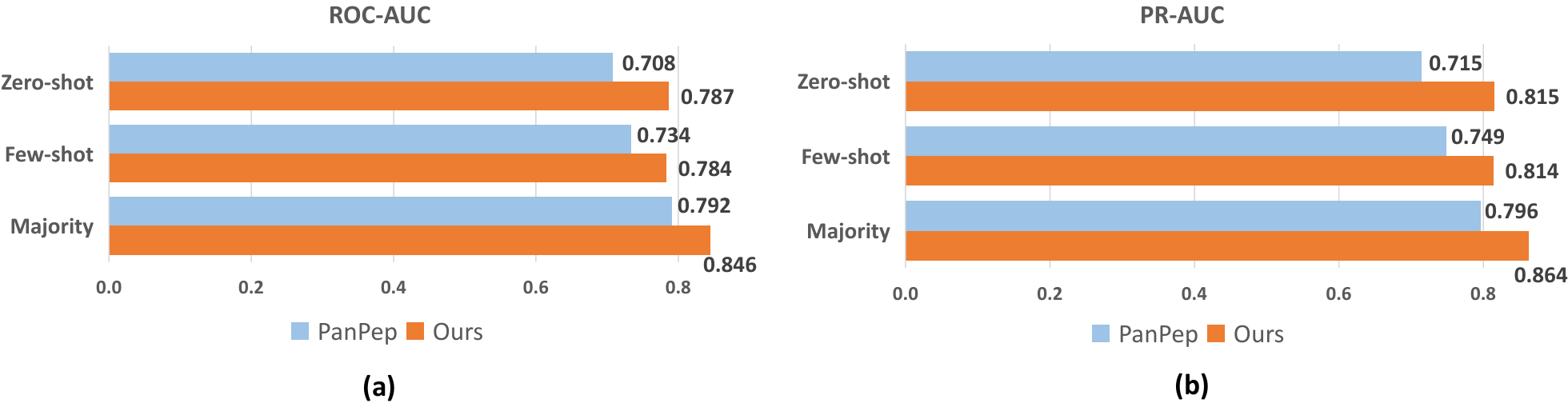}
    \vspace{-0.5em}
    \caption{Comparison to SOTA model (PanPep) of ROC-AUC and PR-AUC performances in the different settings and datasets.}
    \label{fig:diff_setting}
    \vspace{-0.5em}
\end{figure*}

\subsection{Setups}

\textbf{Datasets.} 
Following PanPep\cite{gao2023pan}, we adopt the majority dataset(MajorSet), zero-shot dataset(ZeroSet), and few-shot dataset(FewSet) for primary evaluation. All the binding TCRs are balanced by a controlTCRset, where the controlTCRset contains 60,333,379 non-binding TCRs (negative samples).

\textbf{Evaluation Metrics}.
The evaluation of the proposed models is performed using two important metrics: PR-AUC (Precision-Recall Area Under the Curve) and AU-ROC (Area Under the Receiver Operating Characteristic curve). 

\subsection{DapPep Outperforms Baselines Across Different Settings}

We investigate the generalization ability to predict binding affinity under three different data settings, i.e., majority setting, few-shot setting, and zero-setting, compared with the SOTA model PanPep, as shown in Figure~\ref{fig:diff_setting}. Overall, DapPep has excellent and balanced performances.
For a fair comparison, we use the same evaluation datasets as PanPep.

\textbf{Few-shot Setting.}
We first analyze the comparison in the few-shot setting.
In this setting, the baseline PanPep undergoes a meta-learning approach where the support set for each peptide-specific task in the meta-test dataset is used to finetune the model. In contrast, our trained DapPep does not undergo any continued training or finetuning and is directly validated for generalizability.
As a result, our DapPep achieves an average of 0.784 ROC-AUC and 0.814 PR-AUC in the few settings, while PanPep only achieves an average of 0.734 ROC-AUC and 0.751 PR-AUC.
This analysis shows that DapPep is more adaptive than PanPep in the few-shot setting, and the universal mode has more advantages than the meta-learning here, even without fine-tuning.

\textbf{Zero-shot Setting.}
In the zero-shot setting, the baseline PanPep designs a disentanglement distillation module that extends few-shot learning to zero-shot learning, constructing a mapping between peptide encoding and peptide-specific learners. Here, PanPep has no support set available for model finetuning of newly entered peptides.
The results show that PanPep achieves only 0.708 ROC-AUC and 0.715 PR-AUC, while our DapPep achieves 0.787 ROC-AUC and 0.815 PR-AUC.
This indicates that DapPep can better generalize to unseen peptides.
In comparison with PanPep, our DapPep's performances in the zero-shot setting demonstrate a more pronounced advantage than in the few-shot setting.

\textbf{Majority Setting.}
Finally, we show that both DapPep and PanPep can be easily generalized to a majority setting, where they perform relatively better compared to other settings. Note that here PanPep retrains the model on a large-scale dataset, with different model parameters than those used in the other settings. A MajorSet containing 25 peptide-specific tasks is used to test DapPep and PanPep. Both performed better than expected in MajorSet than in other scenarios. Nevertheless, DapPep still has a bigger advantage in this setting with 0.846 ROC-AUC and 0.864 PR-AUC.


\subsection{DapPep Specializes in Unseen Peptides}


\begin{table}
    \centering
    \begin{tabular}{ccccc}
        \toprule
        \multirow{2}*{\textbf{Models}} & \multicolumn{2}{c}{\textbf{Unseen Peptides}} & \multicolumn{2}{c}{\textbf{TCR Sort}}\\
         & \textbf{ROC-AUC} & \textbf{PR-AUC} & \textbf{ROC-AUC} & \textbf{PR-AUC}\\
        \midrule
        DLpTCR & 0.483 & 0.481 & 0.449 & 0.527\\
        ERGO2 & 0.504 &  0.524 & 0.462 & 0.538\\
        pMTnet & 0.564 &  0.555 & 0.688 & 0.633\\
        PanPep & 0.744 &  0.755 & 0.684 & 0.781\\
        \midrule
        \textbf{DapPep(Ours)} & \textbf{0.816} &  \textbf{0.836} & \textbf{0.835} & \textbf{0.834}\\
        \bottomrule
    \end{tabular}
    \caption{Comparison of ROC-AUC and PR-AUC performances for DapPep and baselines.}
    \label{fig:zero_result}
    \vspace{-1.6em}
\end{table}

Evaluation of the zero-setting is our primary goal. In this test, we compare DapPep with existing tools (PanPep, pMTnet, ERGO2, and DLpTCR) that can predict unseen peptides bound with TCRs. The curated ZeroSet is used as the evaluation set, and the peptides in this dataset are not available in the training set of DapPep and other baseline tools.
As shown in Table~\ref{fig:zero_result}(Unseen Peptides), DapPep significantly outperforms other methods in the zero-setting with a ROC-AUC of 0.816 and a PR-AUC of 0.836 (PanPep: 0.744 and 0.755; pMTnet: 0.563 and 0.555; ERGO2: 0.496 and 0.542; DLpTCR: 0.517 and 0.488).
Compared with SOTA PanPep, our DapPep still surpasses by a large margin (+9.7\% ROC-AUC and +10.7\% PR-AUC), and even surpasses the poor DLpTCR by an amazing +68.9\% ROC-AUC and +73.8\% PR-AUC.
In conclusion, the results demonstrate that DapPep has the potential to predict peptide-specific TCR binding of unseen peptides, showing great promise for exogenous or nascent antigen recognition in various immunological studies and clinical applications.

\subsection{Validation for Potential Clinical Applications}
Adoptive cell transfer is a promising approach for cancer immunotherapy, but its efficiency essentially depends on the enrichment of tumor-specific T cells in the graft \cite{dash2017quantifiable,klebanoff2005sinks}.
We reproduce the results of baselines based on the gastrointestinal cancer dataset\cite{tran2015immunogenicity}, where the dataset study combines next-generation sequencing technology with high-throughput immunological screening to identify tumor-infiltrating lymphocytes from patients with metastatic gastrointestinal cancer, collected from 10 different of neoantigens, and experimentally validated the specific TCRs they bind.
As shown in Table~\ref{fig:zero_result}(TCR Sort), pMTnet, ERGO2, and DlpTC fail in recognizing immunoreactive T cells.
The ROC-AUC of PanPep (0.684) performs poorly, while the PR-AUC (0.781) performs better markedly.
Overall, DapPep performs best in identifying immunoreactive T cells from neoantigens with an ROC-AUC of 0.835 and a PR-AUC of 0.834, which could effectively assist in T cell classification in neoantigen therapy.

\section{Conclusions and Limitations}\label{sec13}
With powerful pre-trained representations, DapPep accomplishes excellent performances on various benchmarks, especially highlighting the ability to generalize to unseen and few peptides, which is meaningful for exogenous or neonatal antigen recognition in computational immunology. Nevertheless, there are potential areas for improvements: 1) Incorporation of additional features or modalities, such as structural information or gene expression data, to enhance the predictive power. 2) Collaboration with experimental biologists and clinicians to validate the predictions in real-world settings and assess their potential for guiding immunotherapeutic interventions.



\bibliographystyle{plain}
\bibliography{my_citation,self_citation}

\end{document}